\documentstyle{mn}

\input epsf

\begin{document}
\journal{Sussex preprint SUSSEX-AST 95/11-1, astro-ph/9511007}
\title[The cluster abundance in flat and open cosmologies]{The cluster 
abundance in flat and open cosmologies}
\author[P.~T.~P.~Viana and A.~R.~Liddle]{Pedro T.~P.~Viana and Andrew 
R.~Liddle\\
Astronomy Centre, University of Sussex, Falmer, Brighton BN1 9QH}
\maketitle
\begin{abstract}
We use the galaxy cluster X-ray temperature distribution function to 
constrain the amplitude of the power spectrum of density inhomogeneities on 
the scale corresponding to clusters. We carry out the analysis for 
critical-density universes, for low-density universes with a cosmological 
constant included to restore spatial flatness and for genuinely open 
universes. That clusters with the same present temperature but different 
formation times have different virial masses is included. We model cluster 
mergers using two completely different approaches, and show that the final 
results from each are extremely similar. We give careful consideration to 
the uncertainties involved, carrying out a Monte Carlo analysis to determine 
the cumulative errors. For critical density our result agrees with previous 
papers, but we believe that the result carries a larger uncertainty. For 
low-density universes, either flat or open, the required amplitude of the 
power spectrum increases as the density is decreased. If all the dark matter 
is taken to be cold, then the cluster abundance constraint remains 
compatible with both galaxy correlation data and the {\it COBE} measurement 
of microwave background anisotropies for any reasonable density.
\end{abstract}

\section{Introduction}

One of the most important constraints that a model of large-scale structure 
must pass is the ability to generate the correct number density of galaxy 
clusters. This is a crucial constraint, because a considerable amount is 
known about clusters from both optical and X-ray measurements. As clusters 
are relatively rare objects, in the standard picture they must form 
from fairly high peaks in the original density field and hence their 
abundance is very sensitive to the normalization of the power spectrum on 
those scales. As a result, the cluster abundance has been cited (White, 
Efstathiou \& Frenk 1993) as one of the strongest pieces of evidence against 
the standard cold dark matter (CDM) model when that model is normalized so 
as to reproduce the microwave background anisotropies as seen by the {\it 
COBE} satellite \cite{Ben96,Ban96,Gor96,Hin96}.

Amongst the many ways in which the CDM model can be salvaged, a popular 
choice has been to reduce the total matter density. The most common context 
within which such models have been studied is in spatially flat cosmologies, 
where a cosmological constant has been introduced to make up the shortfall 
in the matter density \cite{ESM,KGB,KPH}. However, recently attention has 
also been focused on the possibility of a genuinely open cosmological model 
\cite{RPa,Getal2,LLRV}, which possesses somewhat different late-time 
dynamics and also a different {\it COBE} normalization.

In this paper, we re-assess the cluster abundance constraint in CDM models, 
encompassing spatially flat models with both a critical and sub-critical 
matter density and also open universe models. We use the X-ray temperature 
distribution function, taking advantage of a number of extensive 
hydrodynamical cluster simulations which have now been carried out to 
calibrate this to Press--Schechter theory. 

\section{The Power Spectrum}

Except for a brief discussion on {\it COBE} near the end of the paper, we 
shall be interested in scales short enough that even in the open case one 
can consider space to be flat. This allows us to specify the power spectrum 
of the density contrast, following the notation of Liddle \& Lyth 
\shortcite{LL} and Liddle et al. \shortcite{LLRV}, as
\begin{equation}
\label{powerspec}
{\cal P}_{\delta}(k) = \left( \frac{k}{aH} \right)^4 T^2(k) \, 
	\delta_{{\rm H}}^2(k) \, \frac{g^2(\Omega)}{g^2(\Omega_0)} \,,
\end{equation}
where $k$ is the wave number, $a$ is the scale factor, $H = \dot{a}/a$ is 
the Hubble parameter, $\Omega$ is the density parameter, subscript `0' 
indicates the present value, $T(k)$ is the transfer function and the factor 
$g(\Omega)$, defined later, accounts for the rate of growth of density 
perturbations relative to the critical-density case whose growth is given by 
the $(aH)^4$ factor. The quantity $\delta_{{\rm H}}^2(k)$ specifies the 
shape of the primordial spectrum; it is a constant for the 
Harrison--Zel'dovich spectrum, and $\delta_{{\rm H}}^2 \propto k^{n-1}$ for 
a `tilted' spectrum with spectral index $n$. A quantity related to the power 
spectrum is the dispersion $\sigma(R)$ of the density field smoothed on a 
(comoving) scale $R$, defined by 
\begin{equation}
\sigma^2(R) = \int_0^\infty W^2(kR) \, {\cal P}_{\delta}(k) \frac{{{\rm
	d}}k}{k} \,.
\end{equation}
To carry out the smoothing, we shall always use a top-hat window function 
$W(kR)$ defined by
\begin{equation}
W(kR) = 3 \left( \frac{\sin(kR)}{(kR)^3} - \frac{\cos(kR)}{(kR)^2} \right) 
	\,.
\end{equation}

A large galaxy cluster has a mass of around $10^{15} \, {\rm M}_{\sun}$; the 
amount of matter required to make such a cluster would originally (before 
collapse) be contained in a comoving sphere of radius $9.5 
(h/\Omega_0)^{1/3} h^{-1}$ Mpc, where $h$ is the present Hubble constant in 
units of $100 \, {\rm km} \, {\rm s}^{-1} \, {\rm Mpc}^{-1}$. Traditionally, 
the cluster abundance is used to place a constraint on the dispersion of the 
density contrast at the scale $8 h^{-1}$ Mpc, denoted $\sigma_8$, and some 
assumption regarding the shape of the power spectrum is required to shift 
from the scale actually constrained by the observations on to this one. For 
CDM models as considered here, the power spectrum is accurately given by 
Bardeen et al. \shortcite{BBKS} as
\begin{eqnarray}
T_{{\rm CDM}}(q) & = & \frac{\ln \left(1+2.34q \right)}{2.34q} \times  
	\\ \nonumber 
& & \hspace*{-1.5cm}
	\left[1+3.89q+(16.1q)^2+(5.46q)^3+(6.71q)^4\right]^{-1/4} \,,
\end{eqnarray}
with $q = k/h\Gamma$, where the `shape parameter' $\Gamma$ is defined as 
\cite{SUG95}
\begin{equation}
\label{gamma}
\Gamma = \Omega_0 h \exp (-\Omega_{{\rm B}}-\Omega_{{\rm B}}/\Omega_0) \,,
\end{equation}
with $\Omega_{{\rm B}}$ the baryon density. This fit to $\Gamma$ is good 
for both flat and open universes. However, we have no particular need for 
it, as 
we can simply specify models by $\Gamma$. If the spectral index $n$ equals 
$1$, then, provided that $\Gamma$ is chosen in the range $\Gamma = 
0.230^{+0.042}_{-0.034}$ (these limits being 95 per cent confidence), a 
reasonable fit to the shape of the galaxy correlation function is obtained. 
This allowed interval for $\Gamma$ was calculated using the data points in 
table 1 of Peacock \& Dodds \shortcite{PD}, omitting the four corresponding 
to the smallest scales, as at these scales the working assumption of a 
linear bias used in their calculation seems to break down (Peacock 1996). 
Because galaxy correlations are presumed to be biased relative to the mass, 
this fit tells us nothing about the overall normalization of the matter 
power spectrum. 

It is simplest to use an approximation to the true shape of $\sigma(R)$ in 
the vicinity of $8 h^{-1}$ Mpc. White et al. \shortcite{WEF} use a power-law 
fit where the exponent is related to $\Gamma$. Because, particularly for low 
densities, we require accuracy over a greater range of scales, we use the 
more accurate fit
\begin{equation}
\label{sig}
\sigma(R)=\sigma_{8} \left(\frac{R}{8 h^{-1} \; {\rm
	Mpc}} \right)^{-\gamma(R)} \,,
\end{equation}
where for the CDM spectra we adopt the form
\begin{equation}
\label{del}
\gamma(R)=(0.3 \Gamma + 0.2)\left[2.92 + \log \left(\frac{R}{8 h^{-1}
	\; {\rm Mpc}} \right) \right] \,.
\end{equation}
If one chooses a tilted primordial spectrum, $n \neq 1$, then the 
best-fitting $\Gamma$ is changed, but the actual shape of the power spectrum 
is still obliged to fit the shape of the galaxy correlation function and so 
the results that we obtain will not depend significantly on $n$; indeed, 
even the $\Gamma$ dependence will prove unimportant provided that it is 
restricted to lie within the favoured range. 

In a critical-density universe, clusters are anticipated to have formed very 
recently. As we shall see, in low-density universes clusters will have 
formed 
much earlier, and we shall need to take into account the redshift dependence 
of the power spectrum. In CDM universes, the shape of the power spectrum is 
unchanged at low redshift, so we need only consider the redshift dependence 
at a single scale which we take to be $8 h^{-1}$ Mpc. This dependence is 
different depending on whether we have an open model (which we label `OCDM') 
or a flat model (which we label `$\Lambda$CDM'). Following Carroll, Press \& 
Turner \shortcite{CPT} we introduce a growth suppression factor $g(\Omega)$, 
as used in equation (\ref{powerspec}). This gives the total suppression of 
growth of the dispersion relative to that of a critical-density universe, 
and is accurately parametrized by
\begin{eqnarray}
\label{supp}
g(\Omega) & = & \frac{5}{2} \Omega \left[ 1 + \frac{\Omega}{2} + 
	\Omega^{4/7} \right]^{-1} \quad \quad \quad \quad \quad 
	{\rm (OCDM)} \,; \\
\label{suppa}
g(\Omega) & = & \frac{5}{2} \Omega \left[ \frac{1}{70} +
	\frac{209\Omega}{140} -\frac{\Omega^2}{140} + \Omega^{4/7}
	\right]^{-1} \, {\rm (}\Lambda{\rm CDM)} .
\end{eqnarray}
These formulae can be applied at any value of $\Omega$. For a 
matter-dominated low-density universe, the redshift dependence of $\Omega$ 
is given by
\begin{eqnarray}
\label{omo}
\Omega(z) & = & \Omega_0 \, \frac{1+z}{1+\Omega_0 z}  
	\quad \quad \quad \quad \quad \quad \quad {\rm (OCDM)} \,; \\
\label{omf}
\Omega(z) & = & \Omega_0 \, \frac{(1+z)^3}{1-\Omega_0 +(1+z)^3 \Omega_0}
	\quad \quad {\rm (}\Lambda{\rm CDM)} \,.
\end{eqnarray}
Since the growth law in a critical-density universe is $\sigma_8(z) \propto 
(1+z)^{-1}$, the redshift dependence for arbitrary $\Omega_0$ is therefore
\begin{equation}
\label{growth}
\sigma_8(z) = \sigma_8(0) \, \frac{g(\Omega(z))}{g(\Omega_0)} \, 
\frac{1}{1+z} \,,
\end{equation}
using the appropriate formulae for $g(\Omega)$ and $\Omega(z)$ depending on
whether the universe is open or flat.

\section{Press--Schechter Theory}

\subsection{Number densities}

The most accurate way of assessing the cluster abundance is via numerical 
simulation. However, there is an excellent analytic alternative which is the 
Press--Schechter theory \cite{PS,BCEK}. For the case of a critical-density 
universe, this has been extensively tested against $N$-body simulations and 
found to fare extremely well \cite{LC}. Less attention has been directed 
towards testing the theory in open universes, but those studies that do 
exist suggest that it continues to work well, as one might expect since its 
derivation relies only on statistical arguments.

Press--Schechter theory states that the fraction of material residing in 
gravitationally bound systems larger than some given mass $M$ is given by 
the fraction of space in which the {\em linearly evolved} density field, 
smoothed on that mass scale, exceeds some threshold $\delta_{{\rm c}}$. 
Originally the choice of $\delta_{{\rm c}}$ was motivated via the spherical 
collapse model, but it should now be calibrated via $N$-body simulations. 
The appropriate formula is
\begin{equation}
\label{PS}
\frac{\Omega(>M(R),z)}{\Omega(z)} = {\rm erfc} \left( \frac{\delta_{{\rm 
c}}}{\sqrt{2} \, \sigma(R,z)} \right) \,,
\end{equation}
where `erfc' is the complementary error function. Here $R$ is the comoving 
linear scale associated with $M$, $R^3 = 3M/4\pi\rho_{{\rm b}}$, where 
$\rho_{{\rm b}}$ is the {\em comoving} background density which is constant 
during matter domination. In equation (\ref{PS}), the right hand side has 
been multiplied by a factor two to allow the material in underdense regions 
to participate; in original treatments this seemed very ad hoc but it has 
now been more or less well justified \cite{PH,BCEK}, and in any case it is 
incorporated in the $N$-body calibration. The required value of 
$\delta_{{\rm c}}$ is quite dependent on the choice of smoothing window used 
to obtain the dispersion \cite{LC}. We shall use a top-hat window function. 
Were one to use a Gaussian window the threshold would be significantly lower 
(or alternatively the associated mass could be increased as discussed by 
Lacey \& Cole 1994); claims in the literature of a large uncertainty in the 
Press--Schechter calculation are often due to quoting thresholds for 
different types of smoothing and one should always be careful to specify 
which is being used.

For spherical collapse, the appropriate threshold is $\delta_{{\rm c}} = 1.7 
\pm 0.1$. In the generic non-spherical situation, one must be careful to 
specify what is meant by collapse; if one considers collapse along all three 
axes the threshold is higher, whereas collapse along the first axis (pancake 
formation) or the first two axes (filament formation) corresponds to a lower 
threshold \cite{Mon}. Since large clusters are relatively rare, one can 
assume that shear did not play an important role and hence their collapse 
can be considered to have occurred spherically \cite{Ber}.

One might expect $\delta_{{\rm c}}$ to depend on the background cosmology. 
However, this seems not to be the case --- Lilje \shortcite{lilje}, Lacey \& 
Cole \shortcite{LC93} and Colafrancesco \& Vittorio \shortcite{CV} found 
that, at least for any type of collapse where $\delta_{{\rm c}} \leq 1.7$ in 
a flat universe, the value of $\delta_{{\rm c}}$ varies at most by $5$ per 
cent when one goes from a critical-density universe to one with 
$\Omega_{0} = 0.1$. Since at earlier epochs low-density universes become 
closer to critical-density ones, the variation will be even less as one goes 
to higher redshifts. 

Equation (\ref{PS}) gives the total amount of material in structures above a 
given mass. However, we shall be interested in the number density of objects 
within a given mass range. The comoving number density of clusters in a 
mass interval ${\rm d}M$ about $M$ at a redshift $z$ is obtained by 
differentiating equation (\ref{PS}) with respect to the mass and multiplying 
it by $\rho_{{\rm b}}/M$. This gives
\begin{eqnarray}
\label{mfa}
n(M,z) \, {\rm d}M = & & \\ \nonumber
& & \hspace*{-2cm} -\sqrt{\frac{2}{\pi}} \frac{\rho_{{\rm b}}}{M} \,
	\frac{\delta_{{\rm c}}}{\sigma^2(R,z)} \frac{{\rm 
d}\sigma(R,z)}{{\rm
	d}M} \exp \left[- \frac{\delta_{{\rm c}}^2}{2\sigma^2(R,z)}
	\right] {\rm d}M \,.
\end{eqnarray}

Following White et al. \shortcite{WEF}, we can simplify this for $R$ in the 
vicinity of $8h^{-1}$ Mpc by using our analytic approximation to the power 
spectrum, equation (\ref{sig}), to calculate the derivative in equation 
(\ref{mfa}). This gives
\begin{eqnarray}
\label{mf}
n(M,z) \, {\rm d}M = & & \\ \nonumber
& & \hspace*{-2cm} \sqrt{\frac{2}{\pi}}\frac{\rho_{{\rm b}}}{M^2} 
	\frac{2.92(0.3\Gamma + 	0.2) \delta_{{\rm c}}}{3\sigma(R,z)} \exp
	\left[ 
	- \frac{\delta_{{\rm c}}^{2}}{2\sigma^{2}(R,z)} \right] {\rm d}M.
\end{eqnarray}

\subsection{Formation redshifts}

We shall be interested in the formation redshifts of the clusters that we 
see at the present epoch. The literature contains two very different ways of 
obtaining this (Lacey \& Cole 1993, 1994; Sasaki 1994), which each have 
their advantages and disadvantages. In this paper we shall consider both.

Sasaki \shortcite{Sa} showed through simple physical and mathematical 
arguments that the comoving number density of clusters with mass in a range 
${\rm d}M$ about $M$, which virialize in an interval ${\rm d}z$ about some 
redshift $z$ and survive until the present without merging with other 
systems, is given by
\begin{eqnarray}
N(M,z) \, {\rm d}M \, {\rm d}z = & & \\ \nonumber
& & \hspace*{-2cm} \left[ - \frac{\delta_{{\rm c}}^2}{\sigma^2(R,z)} 
	\frac{n(M,z)}{\sigma(R,z)} \frac{{\rm d}\sigma(R,z)}{{\rm d}z}
	\right]\frac{\sigma(R,z)}{\sigma(R,0)} \, {\rm d}M \, {\rm d}z \,.
\end{eqnarray}
The redshift independence of the shape of the CDM power spectrum allows 
us to write
\begin{eqnarray}
\label{ncom}
N(M,z) \, {\rm d}M \, {\rm d}z = & & \\ \nonumber
& & \hspace*{-2cm} \left[ - \frac{\delta_{{\rm c}}^2}{\sigma^2(R,z)} 
	\frac{n(M,z)}{\sigma_8(z)} \frac{{\rm d}\sigma_8(z)}{{\rm d}z}
	\right]\frac{\sigma_8(z)}{\sigma_8(0)} \, {\rm d}M \, {\rm d}z \,,
\end{eqnarray}
where $\sigma_{8}(z)$ and ${\rm d}\sigma_{8}(z)/{\rm d}z$ are calculated 
using equation (\ref{growth}). In equation (\ref{ncom}), the expression 
within the square brackets gives the formation rate of clusters with mass 
$M$ at redshift $z$, whereas the factor outside gives the probability of 
these clusters surviving until the present. The approximation leading to 
this equation was to assume that the efficiency of the destruction rate of 
clusters through mergers has no characteristic mass scale, so that merging 
proceeds in a self-similar way along the entire mass range. Though we do not 
expect this to happen for physically motivated power spectra, such as the 
ones under consideration, it should be a fairly good approximation across a 
limited range of scales. In order to be consistent with the Press--Schechter 
formalism, it then turns out that the efficiency of the destruction rate 
must be only a function of redshift. Blain \& Longair (1993a,b) also worked 
within the Press--Schechter framework, instead assuming various physically 
reasonable merger probability distributions, and they found results 
numerically similar to Sasaki's. 

An alternative approach is that of Lacey \& Cole (1993, 1994), who 
attempted to construct a merging history for dark matter haloes based 
on the random walk trajectories technique. This approach is much closer to 
physical reality than that of Sasaki, but necessarily much more complicated.
They considered two alternative possible techniques for computing the 
distribution of halo formation times, one based on an analytical counting 
method and the other based on Monte Carlo generated merging histories; a 
comparison they made with results from $N$-body simulations shows clearly 
that the first method is better, providing a good fit to the $N$-body 
results. In general, the calculation of the distribution of halo formation 
times using the analytical counting method has to be performed numerically, 
but for a white noise power spectrum, $n=0$, an analytic expression is 
available. Usefully, it turns out that the distribution of halo formation 
times is almost independent of the shape of the power spectrum, so one can 
use the analytical expression for $n=0$ as a good approximation to the 
numerical calculation for any $n$. 

A drawback of the Lacey \& Cole approach is that one has to choose when to 
consider a given dark matter halo to have formed, since the last 
infinitesimal amount of mass is still being accreted at the present. That 
is, at what fraction of its final mass is a halo to be considered to have 
effectively formed? Bearing in mind that different properties of a cluster 
may be affected to different degrees by increasing the cluster mass, clearly 
the definition of when a cluster has effectively formed depends on the 
particular cluster property under study. A cluster will have effectively 
formed if the property barely changes until the cluster reaches its final 
mass. According to Lacey \& Cole (1993, 1994), the probability that a 
cluster with present virial mass $M$ would have formed at some redshift $z$ 
is then given by
\begin{equation}
\label{pzlc}
p(z) = p(w(z)) \frac{{\rm d}w(z)}{{\rm d}z}\,,
\end{equation}
where
\begin{eqnarray}
\label{pwlc}
p(w(z)) = 2 \, w(z) \left( f^{-1}-1 \right) \, {\rm erfc} 
	\left(\frac{w(z)}{\sqrt{2}}\right) - & & \\ \nonumber
& & \hspace*{-5.0 cm} \sqrt{\frac{2}{\pi}} 
	\left( f^{-1}-2 \right) \exp{\left(-\frac{w^{2}(z)}{2}\right)} \,,
\end{eqnarray} 
and
\begin{equation}
w(z) = \frac{\delta_{{\rm c}}\left(\sigma(M,0)/\sigma(M,z) - 1 \right)}
	{\sqrt{\sigma^2 (fM,0) - \sigma^2 (M,0)}}\,,
\end{equation}
with $f$ the fraction of the cluster mass assembled by redshift $z$. As 
the shape of the CDM power spectrum is redshift independent we have 
$\sigma(M,0)/\sigma(M,z) = \sigma_8(0)/\sigma_8(z)$, where $\sigma_8(z)$ 
is given by equation (\ref{growth}). Expression (\ref{pwlc}) was obtained 
for a power-law spectrum with index $n=0$, but, as mentioned before, 
numerical results show that $p(w(z))$ depends only very weakly on $n$ 
(Lacey \& Cole 1993) so we will use it even though, at the scales we are 
interested in, the spectral index is closer to $n = -2$.

\section{The Cluster Abundance}

The first attempts to constrain the power spectrum via the cluster abundance 
were made by Evrard \shortcite{E89} and then by Henry \& Arnaud 
\shortcite{HA}. They both considered only the critical-density case, Evrard 
using the velocity dispersion to determine the mass via the virial theorem 
while Henry \& Arnaud used the X-ray temperature. Both obtained very similar 
results, though the quoted errors of the latter were much smaller. Other 
pre-{\it COBE} analyses were made by Bond \& Myers \shortcite{BM}, by 
Bahcall \& Cen (1992, 1993), by Lilje \shortcite{lilje} who considered 
low-density flat models and by Oukbir \& Blanchard \shortcite{OB} who 
discussed the open case. Subsequent to the {\it COBE} observations, White et 
al. \shortcite{WEF} carried out an analysis with extra input from $N$-body 
simulations, obtaining a similar result again to Evrard \shortcite{E89} and 
Henry \& Arnaud \shortcite{HA} in the case of critical density and extending 
the calculation to the case of flat universes with a low matter density. 
Bartlett \& Silk \shortcite{BS} tested a variety of flat-space models 
against the data, using the then current first year {\it COBE} normalization 
which is some way below that currently recommended \cite{Gor96}. Balland \& 
Blanchard \shortcite{BB} considered the hot dark matter case. Recently, 
Liddle et al. \shortcite{LLRV} briefly described a new calculation in the 
case of an open universe. That calculation extended the type of analysis 
made earlier, by attempting to take into account that clusters with equal 
mass which virialize at different redshifts have distinct properties, such 
as velocity dispersion and X-ray temperature, at the present. As well as 
providing a more detailed account of that open universe calculation, in this 
paper we carry out a similar analysis for flat universes. 

When applied to rare objects, the number density predicted by the 
Press--Schechter theory is extremely sensitive to the dispersion 
$\sigma(R)$. This is of great advantage, because it means that even if the 
number density is not well known the error this induces in estimating 
$\sigma_8$ is small. Much more crucial in this application is to establish 
estimates of the cluster masses as accurately as possible. There are 
presently three methods in use for mass estimation. The cluster velocity 
dispersion provides one such means. Unfortunately, cluster catalogues 
assembled from optical data are prone to contamination from foreground and 
background galaxies mis-identified as part of the cluster, and may also be 
affected by projection effects and by the possibility of velocity bias. An 
alternative means of cluster identification is via X-ray emission from the 
gas resting in a deep potential well. Since X-rays are produced only in 
clusters, there are no problems with foreground and background contamination 
(unless two clusters happen to lie on top of one another in projection, and 
even then discrimination may be possible as luminosity goes as a steep power 
of the cluster richness \cite{HA}). The final method is via weak lensing of 
background galaxies \cite{KS}. This promises to be a very interesting 
technique for the future, but at present has not been widely applied. 
Consequently, we choose to adopt the X-ray data.

The observed number density of clusters per unit temperature, $n(k_{{\rm 
B}}T)$, at $z=0$ has been determined by Edge et al. \shortcite{ESFA} and by 
Henry \& Arnaud \shortcite{HA}. These are in good agreement; we shall use 
the latter\footnote{Eke, Cole \& Frenk \shortcite{ECF} have recently pointed 
out two errors in the derivation by Henry \& Arnaud, which fortunately 
almost exactly cancel each other out.}. We shall concentrate on large 
clusters by considering those with mean X-ray temperature 7 keV; those are 
observed to have a present number density per unit temperature interval of
\begin{equation}
\label{nd}
n(7 \; {\rm keV} , 0) = 2.0_{-1.0}^{+2.0} \times10^{-7}
	h^{3}  \; {\rm Mpc}^{-3} \; {\rm keV}^{-1} \,.
\end{equation}

In order to apply the Press--Schechter formalism, one needs to relate the 
X-ray temperature to the virial mass. As we shall see, in the case of a 
critical-density universe one expects that all clusters formed fairly 
recently and there is more or less a one-to-one correspondence between a 
given temperature and a given virial mass. In low-density universes, 
clusters can form much earlier, and the clusters we see today of a given 
temperature would have formed at a range of different redshifts. Lilje 
\shortcite{lilje} (see also Hanami 1993) has demonstrated that clusters of 
the same present temperature, but different formation times, will in general 
have different virial masses in accordance with the scaling relation 
\begin{eqnarray}
\label{mvprop}
M_{{\rm v}} \propto \Omega_{0}^{-1/2} \, \chi^{-1/2} \,
\left(2\frac{r_{{\rm v}}}{r_{{\rm m}}}\right)^{3/2} \times & & \\ \nonumber
& & \hspace*{-4.0 cm}\left[1-\eta\left(\frac{r_{{\rm v}}}{r_{{\rm m}}}
	\right)^{3}\right]^{-3/2} \, (1+z_{{\rm c}})^{-3/2} \, 
	(k_{{\rm B}}T)^{3/2}h^{-1} \,,
\end{eqnarray}
where
\begin{equation}
\chi = \left(\frac{4}{3\pi}\right)^{2}\xi\,,
\end{equation}
\begin{equation}
\eta = 
2\left(\frac{4}{3\pi}\right)^{2} \left(\frac{\lambda_{0}}{\Omega_{0}}
	\right) \chi^{-1}(1+z_{{\rm m}})^{-3}\,,
\end{equation}
with $\xi$ the ratio between the cluster and background densities at 
turnaround, and
\begin{equation}
\frac{r_{{\rm v}}}{r_{{\rm m}}} = \frac{1-\eta/2}{2+\epsilon-\eta/2}\,,
\end{equation}
where $r_{{\rm m}}$ and $r_{{\rm v}}$ are the radii of turnaround and 
virialization respectively. The parameter $\epsilon$ represents the 
difference in the total energy of a cluster from the one obtained by 
assuming the cluster to be an ideal virialized system collapsed from a 
top-hat perturbation. The two main processes by which such a difference can 
be introduced are radiative cooling and dynamical relaxation due to the 
presence of substructure. In both cases $\epsilon>0$ and there is a loss of 
energy by the cluster, thus making it more compact. Whilst the first is 
generally regarded as unimportant due to the fact that the cooling time of a 
typical rich cluster is larger then the age of the universe, the second 
could have a significant impact on the cluster final state. However, because 
the discussion on the importance of this process is still wide open, we 
choose to use $\epsilon=0$. In the above expressions $\lambda_{0} \equiv 
\Lambda/3H_{0}^{2}$, so $\lambda_{0} = 1 - \Omega_{0}$ in a flat universe. 
Also, $z_{{\rm c}}$ and $z_{{\rm m}}$ are respectively the redshifts of 
cluster collapse and turnaround; they are related by the fact that 
the time of collapse and virialization $t_{{\rm c}}$ is twice the time of 
turnaround $t_{{\rm m}}$, as the expansion and subsequent collapse of a 
cluster are symmetric about the time of turnaround. We shall need to 
calculate $z_{{\rm m}}$ given $z_{{\rm c}}$. The redshift--time relation for 
open models is \cite{KT} 
\begin{eqnarray}
t = H_{0}^{-1}\frac{\Omega_{0}}{2(1-\Omega_{0})^{3/2}} \times & & \\ 
\nonumber
& & \hspace*{-4.0 cm}\left[\frac{2(1-\Omega_{0})^{1/2}(\Omega_{0}z+1)^{1/2}}
{\Omega_{0}(1+z)}-{\rm cosh}^{-1}\left(\frac{\Omega_{0}z-\Omega_{0}+2}
{\Omega_{0}z+\Omega_{0}}\right)\right]\,,
\end{eqnarray}
and for flat models is \cite{CT} 
\begin{eqnarray}
t = H_{0}^{-1}\frac{2}{3\lambda_{0}^{1/2}} \times & & \\ \nonumber 
& & \hspace*{-2.5 cm}{\rm ln}\left\{\frac{[\lambda_{0}(1+z)^{-3}]^{1/2}+
[\lambda_{0}(1+z)^{-3}+\Omega_{0}]^{1/2}}{\Omega_{0}^{1/2}}\right\}\,.
\end{eqnarray}
In the limiting case of a critical-density model we have 
\begin{equation}
t = \frac{2}{3}H_{0}^{-1}(1+z)^{-3/2}\,.
\end{equation}
As we know that $t_{{\rm m}}=t_{{\rm c}}/2$ we can therefore calculate 
$z_{{\rm m}}$ from the above equations through an iterative procedure.

The parameter $\xi$ can be obtained by solving the equation of motion of the 
outermost mass shell of a cluster, and using the fact that at turnaround 
this shell has zero velocity. Following Hanami \shortcite{H} (see also 
Martel 1991) we then have
\begin{equation}
\label{xio}
H_{0}t_{{\rm m}} = \frac{(1+z_{{\rm m}})^{-3/2}}{(\Omega_{0}\xi)^{1/2}}
	\int_{0}^{1} \left(\frac{y}{1-y}\right)^{1/2} {\rm d}y \quad
	{\rm (OCDM)} \,;
\end{equation}
\begin{equation}
\label{xif}
H_{0}t_{{\rm m}} = \sqrt{\frac{\zeta}{\lambda_0}}\int_{0}^{1}
	\left[\frac{y} {1-(1+\zeta)y+\zeta y^{3}}\right]^{1/2} \! \! 
	{\rm d}	y \; \; {\rm (}\Lambda{\rm CDM)} ,
\end{equation}
where
\begin{equation}
\zeta = \frac{\lambda_{0}} {\Omega_{0}(1+z_{{\rm m}})^{3}\xi}\,.
\end{equation}
Whilst for open models the integral in equation (\ref{xio}) can be solved 
analytically, yielding 
\begin{equation}
\xi = \frac{\pi^{2}}{4\Omega_{0}(H_{0}t_{{\rm m}})^{2}}(1+z_{{\rm 
m}})^{-3}\,,
\end{equation}
for flat models the integral in equation (\ref{xif}) has to be solved 
numerically. Hanami \shortcite{H} provided a single fitting function which 
proves good for the open case but not nearly as good for the flat case. We 
calculate improved fitting functions by using the trick that any epoch can 
be regarded as the present provided one uses the appropriate value of 
$\Omega(z)$ as given by equation (\ref{omo}) or (\ref{omf}). Taking the 
epoch of turnaround to be the present leaves $\chi$ depending only on the 
value of $\Omega$, and we can then fit a simple function to the numerical 
result. We find 
\begin{equation}
\chi = \Omega^{-f(\Omega)}\,,
\end{equation}
where $\Omega \equiv \Omega(z_{{\rm m}})$, and within an error of 2 per 
cent 
\begin{eqnarray}
f(\Omega) & = & 0.76 - 0.25 \, \Omega \quad \quad {\rm (OCDM)} \,; \\
f(\Omega) & = & 0.73 - 0.23 \, \Omega \quad \quad {\rm (}\Lambda{\rm CDM)} 
\,.
\end{eqnarray}

The scalings in equation (\ref{mvprop}) have been tested by hydrodynamical 
$N$-body simulations in the case of a critical-density model where they have 
been found to hold very well (Navarro, Frenk \& White 1995).

The crucial question is the value of the proportionality constant in 
equation (\ref{mvprop}). We fix this by taking advantage of a set of 
hydrodynamical $N$-body simulations carried out in the critical-density case 
by White et al. \shortcite{WNEF}. They created a catalogue of twelve 
simulated clusters with different temperatures, and found that a cluster 
with present mean X-ray temperature 7.5 keV corresponds to a mass $M_{{\rm 
A}}=(1.10 \pm 0.22)\times10^{15} \, h^{-1} \; {\rm M}_{\sun}$ within the 
Abell radius ($1.5 h^{-1}$ Mpc) of the cluster centre. The quoted error is 
the 1$\sigma$ dispersion within the catalogue. The conversion from an Abell 
radius to the virial radius is then standard, via the result that the 
simulated clusters have a density profile in the outer regions described by 
$\rho_{{\rm c}}(r) \propto r^{-2.4 \pm 0.1}$ \cite{WNEF,ME,NFW}. In a 
critical-density universe the virial radius encloses a density 178 times the 
background density, and hence through a Monte Carlo procedure, where we 
assume the errors in $M_{{\rm A}}$ and in the exponent of $\rho_{{\rm 
c}}(r)$ to be normally distributed, we obtain $M_{{\rm v}}=(1.23 \pm 0.32) 
\times 10^{15} \, h^{-1} \; {\rm M}_{\sun}$ for the virial mass 
corresponding to a 7.5 keV cluster. The remaining uncertainty is the 
virialization redshift of such a cluster, which is estimated \cite{ME,NFW} 
as $z_{{\rm c}} \simeq 0.05 \pm 0.05$. This enables the normalization of 
equation (\ref{mvprop}), and for the particular case of a 7 keV cluster, the 
observation that we are using, one then has a virial mass given by 
\begin{eqnarray}
\label{mv}
M_{{\rm v}} = (1.2 \pm 0.3) \times 10^{15} \, \Omega_{0}^{-1/2} \,
	\chi^{-1/2} \, \left(2 \frac{r_{{\rm v}}}{r_{{\rm m}}} 
	\right)^{3/2} \times & & \\ \nonumber
& & \hspace*{-6 cm}\left[1-\eta 
	\left( \frac{r_{{\rm v}}}{r_{{\rm m}}} \right)^{3} \right]^{-3/2} \,
	(1+z_{{\rm c}})^{-3/2} \, h^{-1} \; {\rm M}_{\sun}\,.
\end{eqnarray}

We shall now compute the present comoving number density of clusters for a 
unit temperature interval about 7 keV in a given cosmology, in order to 
compare it with the observed value. We will do this in two different 
ways, one using the results of Sasaki \shortcite{Sa} and the other those of 
Lacey \& Cole (1993, 1994).

\subsection{The Sasaki method}

According to Sasaki \shortcite{Sa}, the comoving number density of clusters 
that virialize at redshift $z$ with some virial mass $M$, and that then 
survive up to the present without merging with other systems, is given by 
equation (\ref{ncom}). For our application, we wish to consider at 
each redshift the appropriate virial mass which gives rise to a 7 keV 
cluster, via equation (\ref{mv}) with $z_{{\rm c}}=z$. Using the chain rule, 
we can obtain the comoving number density of clusters per temperature 
interval ${\rm d}(k_{{\rm B}}T)$ that virialize at each redshift $z$ and 
survive up to the present such that they have a present mean X-ray 
temperature of 7 keV: 
\begin{eqnarray}
N(k_{{\rm B}}T,z) \, {\rm d}(k_{{\rm B}}T) \, {\rm d}z & =  &  
	\frac{{\rm d}M}{{\rm d}(k_{{\rm B}}T)} 	N(M,z) \, 
	{\rm d}(k_{{\rm B}}T) \, {\rm d}z \nonumber \\ 
 & = & \frac{3}{2} \frac{M}{k_{{\rm B}}T}N(M,z) \, {\rm d}(k_{{\rm B}}T) 
 	\, {\rm d}z \,,
\end{eqnarray}
where the second equality uses equation (\ref{mvprop}). This yields the 
final expression
\begin{eqnarray}
\label{nt}
N(k_{{\rm B}}T,z) \, {\rm d}(k_{{\rm B}}T) \, {\rm d}z = & & \\ \nonumber
& & \hspace*{-2.5 cm} - \frac{3}{2} \frac{M}{k_{{\rm B}}T}
	\frac{\delta_{{\rm c}}^2}{\sigma^2(R,z)} 
	\frac{n(M,z)}{\sigma_8(z=0)} \frac{{\rm d}\sigma_8(z)}{{\rm d}z} 
	\, {\rm d}(k_{{\rm B}}T) \, {\rm d}z \,.
\end{eqnarray}
In order to compute the present comoving number density of clusters per unit 
temperature at 7 keV, one needs to integrate this expression from redshift 
zero up to infinity with the appropriate cosmological information inserted. 
Fig.~1 shows plots of the integrand as a function of redshift, thus 
indicating when the surviving clusters predominantly formed, for a 
critical-density model and for both a flat and an open model with $\Omega_0 
= 0.3$. From the integrated expression, one can find the required $\sigma_8$ 
to obtain the observed number density.

For both open and flat models, one can quote the result in the form of the 
required $\sigma_8$ as a function of $\Omega_0$. The best-fitting value is 
given by
\begin{equation}
\label{final1}
\sigma_8 = 0.60 \, \Omega_0^{-C(\Omega_{0})}\,
\end{equation}
where
\begin{eqnarray}
C(\Omega_0) & = & 0.34 + 0.28 \, \Omega_0 - 0.13 \,\Omega_0^2  \quad \quad
	{\rm (OCDM)} \,, \\
C(\Omega_0) & = & 0.57 - 0.13 \, \Omega_0 + 0.08 \, \Omega_0^2 \quad \quad
	{\rm (}\Lambda{\rm CDM)}  
\end{eqnarray}
are fitting functions representing the changing power-law index of the 
$\Omega_0$ dependence and are accurate within 2 per cent\footnote{Note that, 
although we have made several alterations as compared to the OCDM 
calculation in Liddle et al. \shortcite{LLRV}, the final result is very 
similar.}.

\begin{figure}
\centering
\leavevmode\epsfysize=5.5cm \epsfbox{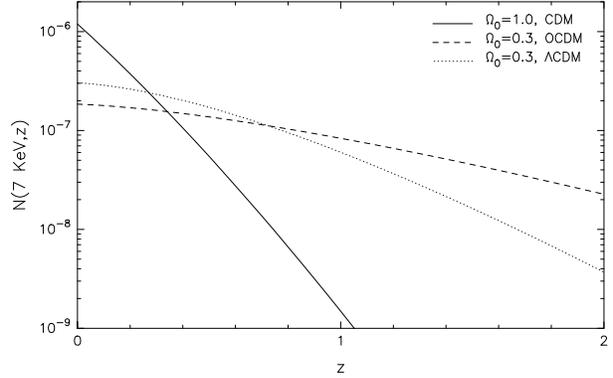}
\caption[figure1]{This shows the differential rate of formation of clusters 
with mean X-ray temperature 7 keV that survive to the present using the 
Sasaki method, where we take ${\rm d}(k_{{\rm B}}T)$ to be the unit 
interval. Curves are shown for three cosmological models: a critical-density 
universe (solid), an open universe with $\Omega_0 = 0.3$ (dashed) and a flat 
model with $\Omega_0 = 0.3$ (dotted). Each curve is normalized to produce 
the same present number density.}
\end{figure}

While the central value is itself of interest, it is vital to know what 
range of values of $\sigma_8$ about this is permitted. The uncertainties 
arise from a variety of sources. Ranking them in order of importance as 
contributions to the overall uncertainty in $\sigma_8$, we have first the 
uncertainties in relating the cluster temperature to its virial mass in 
equation (\ref{mv}) and the uncertainty in the Press--Schechter 
threshold parameter $\delta_{{\rm c}}$, then the uncertainty in the observed 
number density in equation (\ref{nd}), and finally the uncertainty in the 
observational value of $\Gamma$. Since the analysis is both numerical and 
non-linear, we estimate the errors via a Monte Carlo procedure, whereby we 
model the different uncertainties as Gaussians (the observed number density 
is modelled as a Gaussian in its logarithm, as suggested by the error bars). 
Realizations are then drawn from the Gaussian distributions and processed 
through the calculation to determine the required $\sigma_8$; the 
distribution of these is then taken as the uncertainty in $\sigma_8$. While 
this procedure implicitly imagines the errors to be statistical whereas in 
reality they may be predominantly systematic, the fact that there are 
several sources of errors, none of which dominates completely, means that 
this procedure should not be excessively stringent, and indeed our estimate 
of the uncertainty turns out to be rather larger than others that have 
appeared in the literature thus far.

The uncertainty in the observed slope of the power spectrum $\Gamma$ 
is included in the overall uncertainty. We find that, for each $\Omega_0$ 
between 0.1 and 1.0, the distribution of $\sigma_8$ is close to lognormal. 
In the OCDM case, the 95 per cent confidence limits are $+32$ per cent and 
$-24$ per cent, independent of $\Omega_0$ to a good approximation. In the 
$\Lambda$CDM case, the uncertainty becomes larger at low density; a 
satisfactory fit to this increase in uncertainty is to multiply the OCDM 
uncertainty by a factor $\Omega_0^{0.24{\rm log}_{10} \Omega_0}$. For 
example, in a flat universe with $\Omega_0 = 0.3$ the 95 per cent confidence 
limits are increased to $+37$ per cent and $-28$ per cent, and the 
uncertainty climbs rapidly if $\Omega_0$ is further reduced.\footnote{We 
also made an analysis where $\delta_{{\rm c}}$ was shifted to its 95 per 
cent limit and then the remaining errors analysed via the Monte Carlo 
procedure; this corresponds to treating the $\delta_{{\rm c}}$ uncertainty 
as entirely systematic. If one does this, the error bars at 95 per cent 
confidence are increased to $+42$ per cent and $-31$ per cent in the OCDM 
case, the relative increase being the same in the $\Lambda$CDM case.}

\subsection{The Lacey \& Cole method}

We will now use the results from Lacey \& Cole (1993) to obtain an 
alternative estimate of $\sigma_{8}(\Omega_{0})$ for open and flat 
cosmological models. According to equation (\ref{pzlc}), one can estimate 
the fraction of present clusters with virial mass $M$ that formed at a 
redshift $z$. As we want to count only those present clusters with a mean 
X-ray temperature of 7 keV, for each $M$ the corresponding value of $z$ is 
uniquely fixed by equation (\ref{mv}). Therefore the product of the present 
comoving number density of clusters per unit mass with virial mass $M$ 
(given by equation (\ref{mf}) with $z=0$) with the fraction of those 
clusters with a present mean X-ray temperature of 7 keV (given by equation 
(\ref{pzlc}) with $z$ obtained from equation (\ref{mv}) for the given $M$) 
uniquely defines the comoving number density of clusters per unit mass with 
present mean X-ray temperature 7 keV that formed at redshift $z$:
\begin{equation}
\label{lcr}
N(M, z){\rm d}M{\rm d}z = n(M, z=0) \, p(z) \, {\rm d}M{\rm d}z \,.
\end{equation}
Again using the chain rule, the present comoving number density of clusters 
per temperature interval ${\rm d}(k_{{\rm B}}T)$ with a mean X-ray 
temperature of 7 keV that formed at each redshift $z$ is then given by
\begin{equation}
\label{lcrn}
N(k_{{\rm B}}T, z){\rm d}(k_{{\rm B}}T){\rm d}z = 
	\frac{3}{2}\frac{M}{k_{{\rm B}}T}n(M, z=0) \, p(z)
	\, {\rm d}M{\rm d}z \,.
\end{equation}
Once more the present comoving number density of clusters per unit 
temperature at 7 keV is obtained by integrating this expression from 
redshift zero up to infinity.

As we mentioned in the last section, the approach of Lacey \& Cole 
(1993, 1994) has the drawback of requiring one to define a criterion for 
when a cluster is effectively formed in terms of the fraction of the cluster 
final mass assembled at that moment, $f$. In their work Lacey \& Cole 
defined the formation time of a cluster as the moment when half 
of its final mass has been assembled. For our purposes we require the moment 
after which any cluster mass increase leads to only a small change in the 
cluster temperature. To our knowledge the only comparison between the 
evolution of a cluster's mass and its X-ray temperature is that of Navarro 
et al. \shortcite{NFW}. We have already used their results to estimate the 
formation redshift of clusters in a critical-density universe in order to 
normalize equation (\ref{mvprop}), where we took $z_{{\rm c}} = 0.05 \pm 
0.05$. We can read from Navarro et al. \shortcite{NFW} how much mass the 
clusters had assembled within that approximate redshift interval. This gives 
$f = 0.75 \pm 0.15$, where the error is supposed to correspond to a 
2$\sigma$ confidence interval. This seems a physically reasonable result, 
and we shall assume that it remains true in all cosmologies. 

In Fig.~2 we show plots of expression (\ref{lcrn}) as a function of 
redshift, thus indicating when the clusters of temperature 7 keV 
predominantly formed, defined as when 75 per cent of their mass had 
assembled.

Integrating expression (\ref{lcrn}) from redshift zero to infinity and 
comparing the result with the observed number density (\ref{nd}) we once 
more obtain $\sigma_8$ as a function of $\Omega_0$ for both open and flat 
models. Using the Lacey \& Cole approach the best-fitting value is given 
with an error of less than 2 per cent by
\begin{equation}
\label{final2}
\sigma_8 = 0.60 \, \Omega_0^{-C(\Omega_{0})}\,
\end{equation}
with
\begin{eqnarray}
C(\Omega_0) & = & 0.39 + 0.23 \, \Omega_0 - 0.23 \,\Omega_0^2  \quad \quad
	{\rm (OCDM)} \,; \\
C(\Omega_0) & = & 0.61 - 0.19 \, \Omega_0 + 0.02 \, \Omega_0^2 \quad \quad
	{\rm (}\Lambda{\rm CDM)} \,.
\end{eqnarray}

\begin{figure}
\centering
\leavevmode\epsfysize=5.5cm \epsfbox{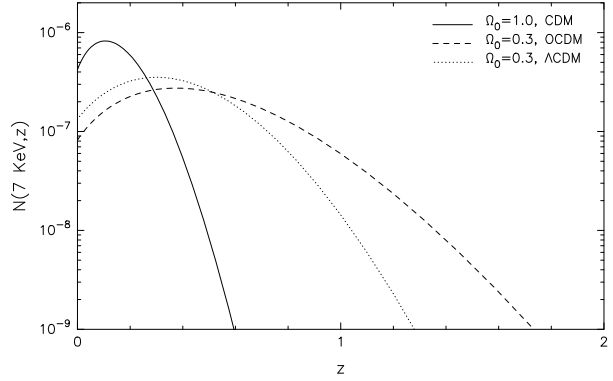}
\caption[figure2]{This shows the differential rate of formation of clusters 
with present mean X-ray temperature 7 keV using the Lacey \& Cole method, 
where we take ${\rm d}(k_{{\rm B}}T)$ to be the unit interval. Curves are 
shown for three cosmological models: a critical-density universe (solid), an 
open universe with $\Omega_0 = 0.3$ (dashed) and a flat model with $\Omega_0 
= 0.3$ (dotted). Each curve is normalized to produce the same present number 
density.}
\end{figure}

The overall error in the value of $\sigma_8$ obtained using the method of 
Lacey \& Cole \shortcite{LC93} comes from the same uncertainties that
we had using the method by Sasaki \shortcite{Sa}, plus the uncertainty in 
the value of the fraction of the cluster mass assembled at the time of 
effective formation. We again estimate the overall error via a Monte Carlo 
procedure, finding that the distribution of $\sigma_8$ is still close to 
lognormal. In both the OCDM and $\Lambda$CDM cases the results obtained 
using the method of Lacey \& Cole \shortcite{LC93} show an increase in 
the overall error in the value of $\sigma_8$ relative to the method of 
Sasaki \shortcite{Sa}. This increase becomes larger at low densities and is 
more important in the OCDM case. The 95 per cent confidence limits for the 
value of $\sigma_8$ calculated using the method of Lacey \& Cole 
\shortcite{LC93} are given to a good approximation by $+32\Omega_0^{0.17{\rm 
log}_{10} \Omega_0}$ per cent and $-24\Omega_0^{0.17{\rm log}_{10} 
\Omega_0}$ per cent in the OCDM case, and $+32\Omega_0^{0.26{\rm log}_{10} 
\Omega_0}$ per cent and $-24\Omega_0^{0.26{\rm log}_{10} \Omega_0}$ per cent 
in the $\Lambda$CDM case. For example, in an open universe with $\Omega_0 = 
0.3$ the 95 per cent confidence limits are $+36$ per cent and $-27$ per 
cent, and in a flat universe, also with $\Omega_0 = 0.3$, the 95 per cent 
confidence limits are $+38$ per cent and $-28$ per cent.

\section{Discussion}

In Fig.~3 we compare the values obtained for $\sigma_8(\Omega_0)$ using the 
two different methods of Sasaki \shortcite{Sa} and Lacey \& Cole 
\shortcite{LC93}. For OCDM the difference is less than 5 per cent down to 
$\Omega=0.2$, increasing to 11 per cent for $\Omega=0.1$, whilst for 
$\Lambda$CDM the difference is less than 3 per cent down to $\Omega=0.2$, 
increasing to 9 per cent for $\Omega=0.1$. The difference between 
the results is much less than the other uncertainties. In order to present a 
definite result, we fit to the mean of the two methods to obtain our final 
result:
\begin{equation}
\sigma_8 = 0.60 \, \Omega_0^{-C(\Omega_{0})}\,
\end{equation}
with
\begin{eqnarray}
C(\Omega_0) & = & 0.36 + 0.31 \, \Omega_0 - 0.28 \,\Omega_0^2  \quad \quad
	{\rm (OCDM)} \,; \\
C(\Omega_0) & = & 0.59 - 0.16 \, \Omega_0 + 0.06 \, \Omega_0^2 \quad \quad
	{\rm (}\Lambda{\rm CDM)} \,.
\end{eqnarray}
Since the difference between the two methods is small, the overall 95 per 
cent confidence limits remain as before. The most conservative assumption 
is to consider those obtained using the method of Lacey \& Cole 
\shortcite{LC93}, which are given by $+32\Omega_0^{0.17{\rm log}_{10} 
\Omega_0}$ per cent and $-24\Omega_0^{0.17{\rm log}_{10} \Omega_0}$ per cent 
in the OCDM case, and $+32\Omega_0^{0.26{\rm log}_{10} \Omega_0}$ per cent 
and $-24\Omega_0^{0.26{\rm log}_{10} \Omega_0}$ per cent in the $\Lambda$CDM 
case. 

\begin{figure}
\centering
\leavevmode\epsfysize=5.5cm \epsfbox{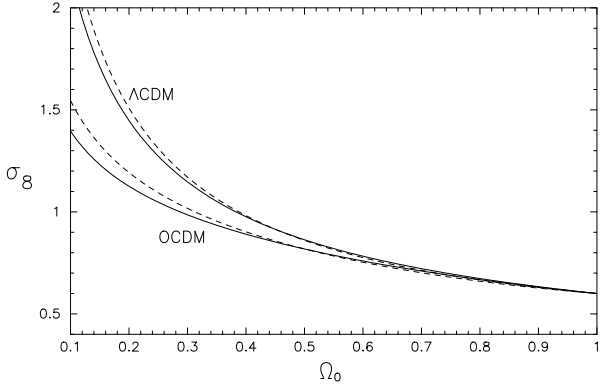}
\caption[figure3]{The dependence of $\sigma_8$ on $\Omega_0$ is shown 
as obtained using the two different methods of Sasaki (solid lines) and 
Lacey \& Cole (dashed lines). The lower two curves are for the open 
case, and the upper two for the flat case.}
\end{figure}

\begin{figure}
\centering
\leavevmode\epsfysize=5.5cm \epsfbox{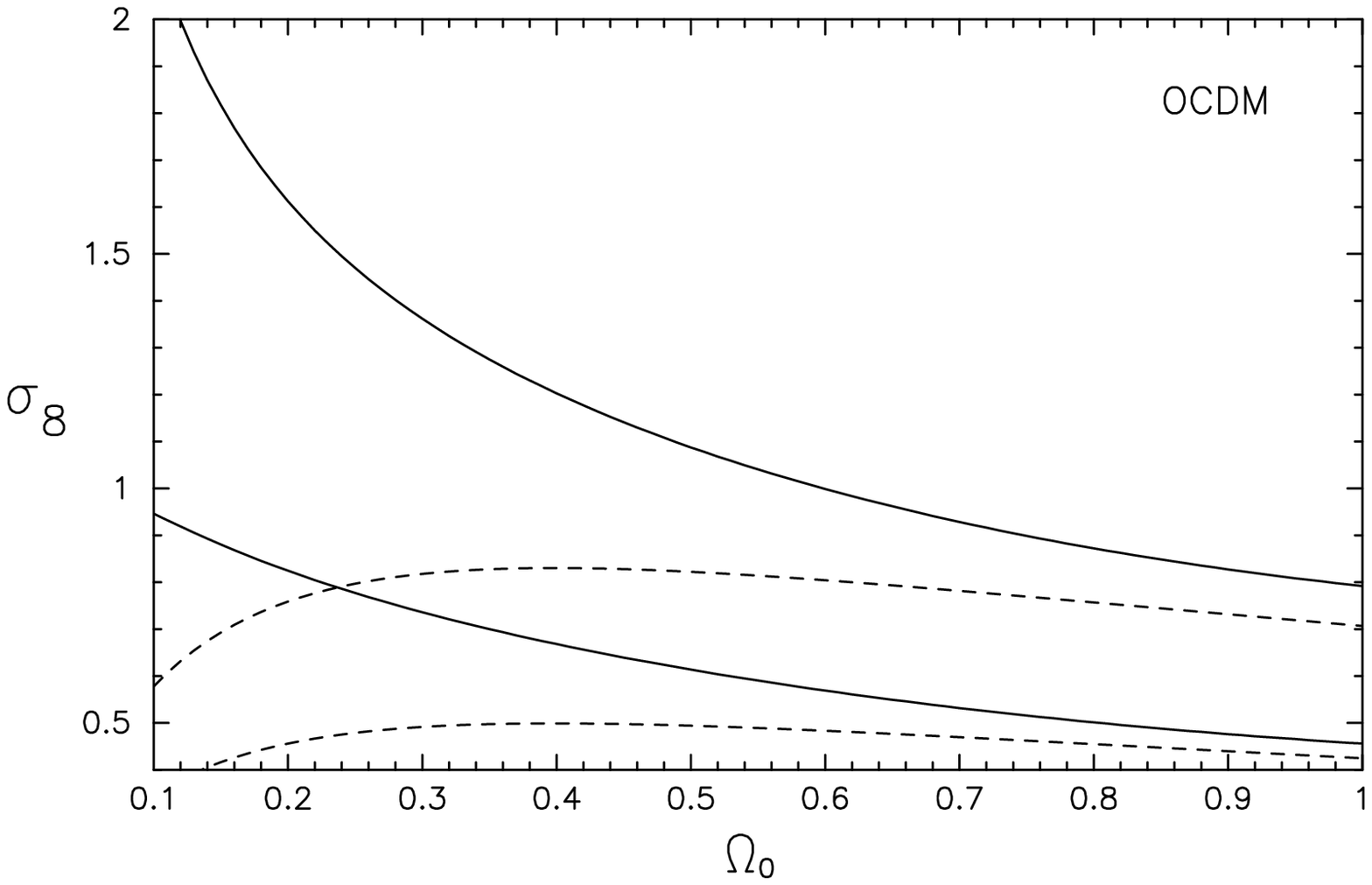}\\
\leavevmode\epsfysize=5.5cm \epsfbox{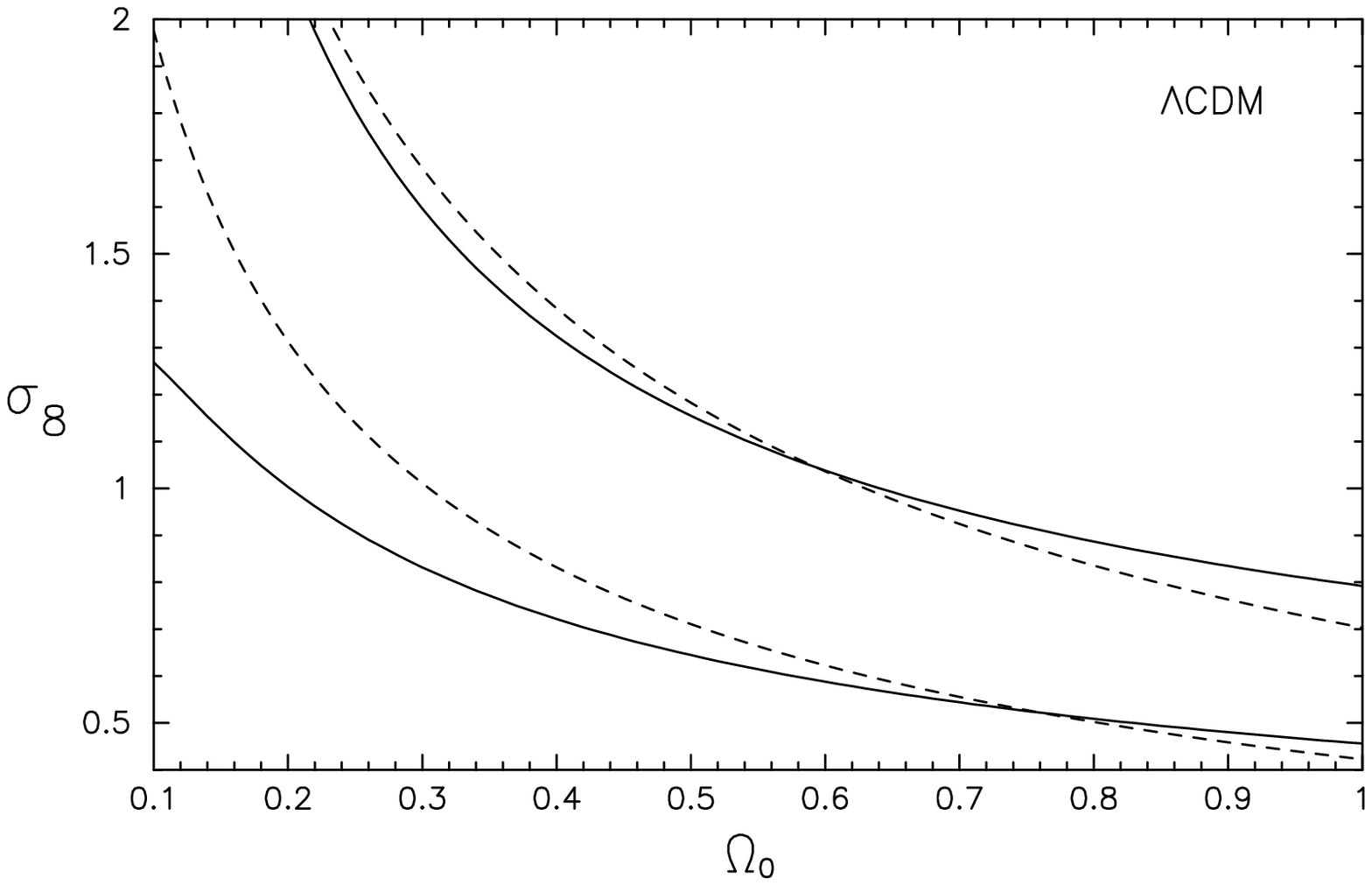}
\caption[figure4]{The cluster constraint (solid) is compared with the {\it 
COBE} normalization (dashed) for the range of $\Gamma$ satisfying the galaxy 
correlation data. The upper figure shows the open case, and the 
lower one the flat case. In each, it is assumed that all the dark matter is 
cold. The upper {\it COBE} line corresponds to $\Gamma = 0.27$ and the lower 
one to $\Gamma = 0.20$. Both cluster and {\it COBE} constraints are at 95 
per cent confidence. The central values are not shown.}
\end{figure}

We do not intend a detailed comparison of these results with other types of 
observation here (see Liddle et al. 1995, 1996a, 1996b), but it is worth 
comparing the results with the normalization from the four-year {\it COBE} 
observations \cite{Ben96,Ban96,Gor96,Hin96} for the case of scale-invariant 
($n=1$) initial density perturbations. These fix the amplitude of 
perturbations on large scales, $\delta_{{\rm H}}^2$ as a function of 
$\Omega_0$, independently of $h$ to an excellent approximation. Fitting 
functions have been calculated by Bunn \& White (in preparation) using 
techniques from White \& Bunn \shortcite{WB}, and are
\begin{eqnarray}
\label{COBEnorm}
10^{10} \delta_{{\rm H}}^2(\Omega_0) & = & 3.80 \, \Omega_0^{-0.70-0.38
	\ln\Omega_{0}} \quad {\rm (OCDM)} \,; \\
10^{10} \delta_{{\rm H}}^2(\Omega_0) & = & 3.76 \, \Omega_0^{-1.57-0.10
	\ln\Omega_{0}} \quad {\rm (}\Lambda{\rm CDM)} \,.
\end{eqnarray}
The fitting functions are accurate within about 3 per cent for $0.2 < 
\Omega_0 < 1.0$. The {\it COBE} normalization error bar on $\delta_{{\rm 
H}}$ is 15 per cent at 2$\sigma$.

Although these values are independent of $h$, and indeed of the nature of 
the dark matter, they pick up a dependence on these when one computes the 
equivalent $\sigma_8$. We shall assume that the dark matter is all cold, so 
that we can use the $\Gamma$ parametrization of the transfer function. The 
uncertainty in the observed value of $\Gamma$ given after equation 
(\ref{gamma}) propagates to the calculated value for $\sigma_8$ when one 
uses the appropriate {\it COBE} normalization for some value of 
$\Omega_{0}$. We will therefore add in quadrature the uncertainty thus 
arising in $\sigma_8$ and the 15 per cent error at 2$\sigma$ which appears 
in $\sigma_8$ due to the {\it COBE} normalization uncertainty. 

The comparison is plotted in Fig.~4. The allowed range of values for 
$\Omega_{0}$ goes down to about 0.20 in the OCDM case. In the $\Lambda$CDM 
case the cluster constraint remains compatible with {\it COBE} and the 
galaxy correlation function shape down to very low $\Omega_0$. However, in 
this latter case there are other reasons for believing that low values are 
not favoured, for example, values of $\Omega_{0}$ below 0.4 require an 
anti-bias for optically selected galaxies due to their very high {\it COBE} 
normalization. This is difficult to understand physically. The anti-bias of 
{\it IRAS} galaxies would be even more pronounced. However, this conclusion 
is driven by {\it COBE} rather than the cluster abundance and can perhaps be 
partly alleviated by tilting the primordial spectrum.

For completeness, in Fig.~5 we show the comoving number density per unit 
temperature interval of clusters with mean X-ray temperature 7 keV that we 
should expect to observe at redshift $z$, for three cosmological 
models normalized to the present central value. The Sasaki and Lacey \& Cole 
methods give very similar results. Note that these are clusters which have 
that temperature at the given redshift; their X-rays would be redshifted on 
the way to us, hence making the apparent temperature smaller by a factor of 
$(1+z)$. As one progresses to a lower redshift, some of the clusters 
will merge and some new ones will form. At present the only complete surveys 
of clusters at high redshifts are flux-limited (e.g. Henry, Jiao \& Gioia 
1994). This poses various problems, the most serious of which is that as the 
cluster selection is then a function of their luminosity, not of their 
temperature, we need to know the temperature--luminosity relationship at the 
appropriate redshift to be able to compare the observations with our 
results. However, this relation is very sensitive to several parameters like 
the cluster baryon fraction and the extent to which energy is injected into 
or removed from the intracluster medium, examples of the former being gas 
stripping from galaxies infalling into the cluster or supernova explosions 
in cluster galaxies, and of the latter being cooling flows. 

\begin{figure}
\centering
\leavevmode\epsfysize=5.5cm \epsfbox{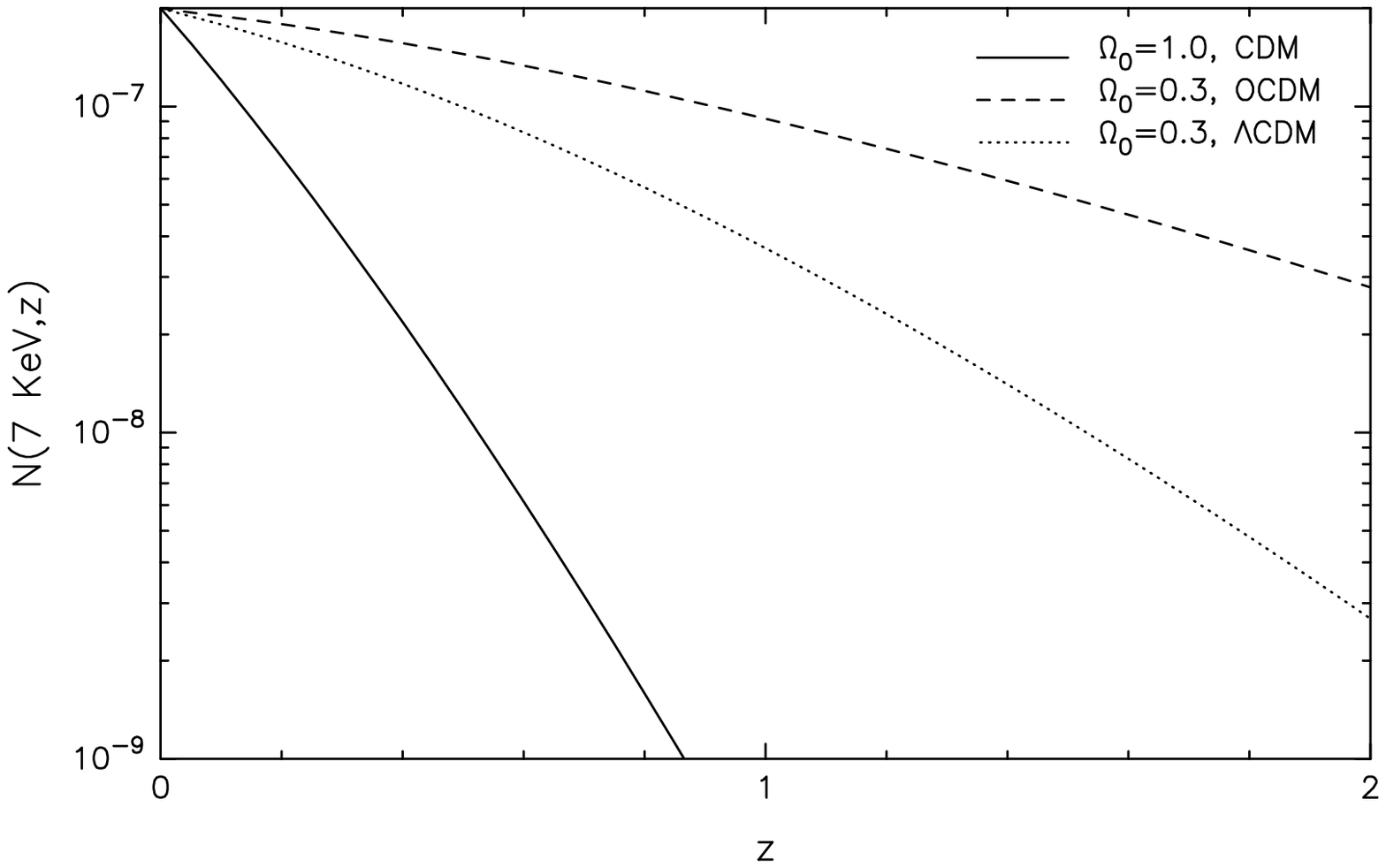}\\
\leavevmode\epsfysize=5.5cm \epsfbox{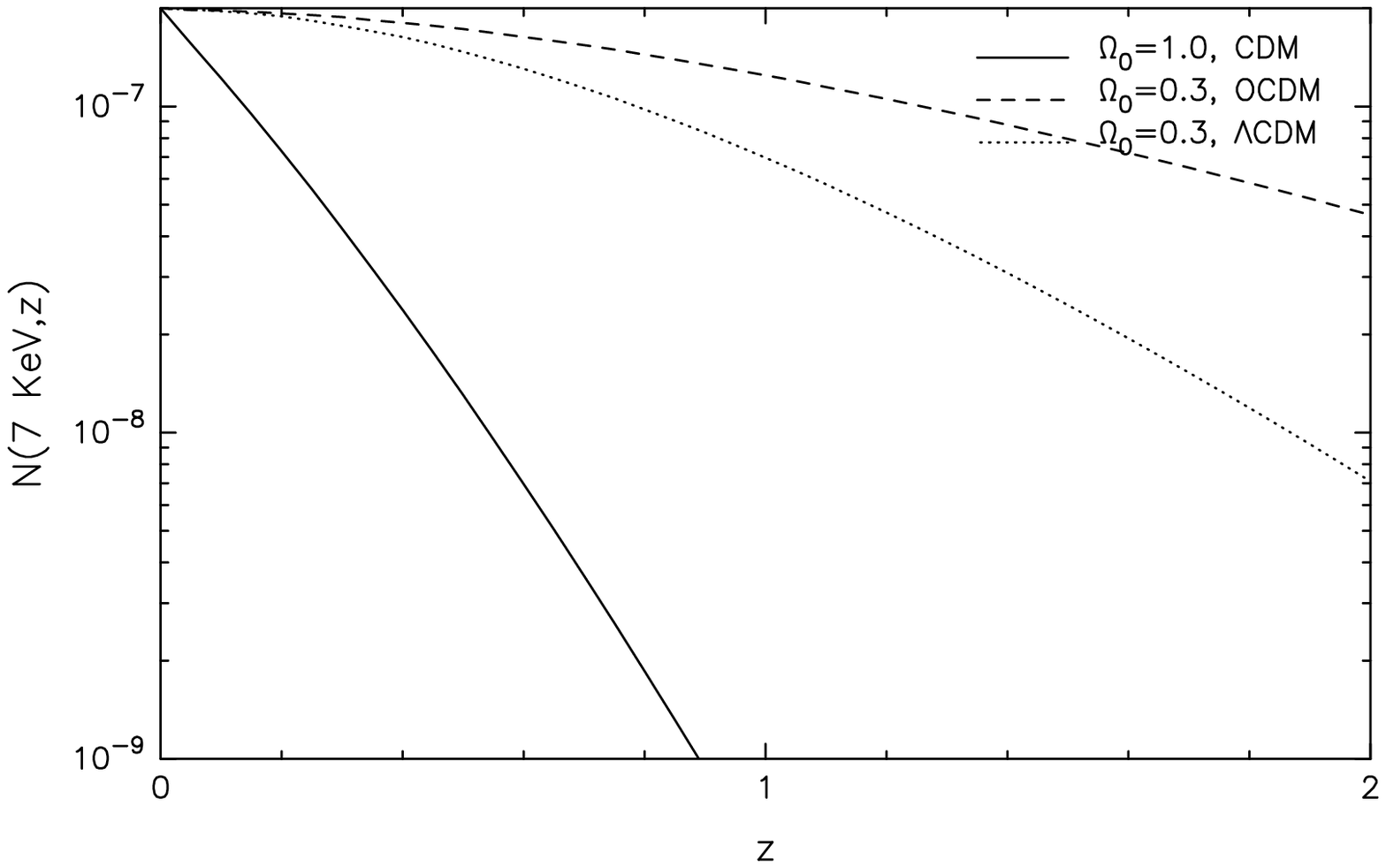}\\
\caption[figure5]{The expected comoving number density of clusters 
for a unit temperature interval about 7 keV at redshift $z$ for three 
cosmological models: a critical-density universe (solid), an open universe 
with $\Omega_0 = 0.3$ (dashed) and a flat model with $\Omega_0 = 0.3$ 
(dotted). Each curve is normalized to produce the same present number 
density. The upper panel shows the result using the Sasaki merger 
calculation, and the lower one that of Lacey \& Cole.}
\end{figure}

To recap on our main results, we have used a new method to compute the power 
spectrum constraint arising from the observed number density of clusters 
with a given X-ray temperature\footnote{After we submitted our paper, a 
preprint appeared by Eke et al. \shortcite{ECF} which overlaps with our 
paper. The principal results are in reasonable agreement.}. We have done 
this for a critical-density universe and for both flat and open low-density 
universes. Although we have assumed that all the dark matter is cold, the 
constraint would not change much \cite{LLSSV} if one went, for example, to a 
cold plus hot dark matter model. In cases where such calculations have been 
done previously, we support the previous results but typically find larger 
accumulated uncertainties.

For the case of pure cold dark matter, we have also compared our results 
with constraints from {\it COBE} and from the galaxy correlation function 
(which constrains the shape parameter $\Gamma$). We found that the cluster 
constraint is compatible with these in both the flat and open cases for any 
reasonable value of the density, failing only in the open case for $\Omega_0 
< 0.20$. However, near critical density the corresponding $h$ will be very 
low (below 0.5), and at the lowest permitted densities it will be very high 
(above 1.0). Further, in the spatially flat case the concordance of these 
constraints at low values of $\Omega_0$ will require optically selected 
galaxies to be anti-biased with respect to the dark matter distribution.
 
\section*{ACKNOWLEDGMENTS}

PTPV is supported by the PRAXIS XXI programme of JNICT (Portugal), and ARL 
by the Royal Society. We are extremely grateful to Cedric Lacey for a 
detailed series of comments leading to significant improvements in this 
paper. ARL acknowledges the hospitality of TAC (Copenhagen) where these 
discussions took place. We are indebted to Ted Bunn and Martin White for 
allowing us to use their four-year {\it COBE} normalization in advance of 
publication. We also thank David Lyth and Frazer Pearce for discussions and 
Jim Bartlett for a helpful referee's report. We acknowledge the use of the 
Starlink computer system at the University of Sussex. 


\bsp
\end{document}